\input{aipcheck}

\newcommand{\lya}{\mbox{${\rm Ly}\alpha$}}
\newcommand{\hI}{\mbox{${\rm H\ I}$}}
\newcommand{\etal}{et al.}
\newcommand{\kms}{\mbox{km\ s${^{-1}}$}}

\documentclass[
  ,draft            % use draft while you are working on the paper
  ,numberedheadings % uncomment this option for numbered sections
  ,letterpaper                 % add further options here if necessary
  ]
  {aipproc}

\layoutstyle{6x9}

\begin{document}

\title{GRB Afterglows as a New ISM and IGM Probe}

\classification{98.70.Rz}
\keywords      {gamma rays: bursts---quasars: absorption lines}

\author{Hsiao-Wen Chen}{
  address={Department of Astronomy \& Astrophysics, University of Chicago,
           Chicago, IL 60637, U.S.A.},
  email={hchen@oddjob.uchicago.edu}
}

\author{Jason X.\ Prochaska}{
  address={UCO/Lick Observatory, University of California at Santa Cruz, 
           Santa Cruz, CA 95064, U.S.A.},
  email={xavier@ucolick.org}
}

\author{Josh S.\ Bloom}{
  address={Department of Astronomy, University of California at Berkeley,
           Berkeley, CA 94720, U.S.A.},
  email={jbloom@astron.berkeley.edu}
}

\begin{abstract}

We summarize results from a study of the metallicity, relative
abundances, gas density, and kinematics of dense media in the host
environment of two {\em Swift} bursts, GRB\,050730 at $z=3.968$ and
GRB\,051111 at $z=1.549$.  Both GRB hosts exhibit strong absorption
features from excited Si$^+$ and Fe$^+$ ions, indicating an extreme
ISM environment that is similar to what is found around massive stars
like luminous blue variables (LBV) and Wolf-Rayet stars.  The extreme
ISM properties have
never been observed in intervening quasar absorption line systems
beyond the local universe.

\end{abstract}

\maketitle

%%%%%%%%%%%%%%%%%%%%%%%%%%%%%%%%%%%%%%%%%%%%
%% MAINMATTER
%%%%%%%%%%%%%%%%%%%%%%%%%%%%%%%%%%%%%%%%%%%%

\section{Unveiling Extreme Star-forming Regions}

Mounting evidence has demonstrated that long-duration gamma-ray bursts
(GRBs) arise in active star-forming regions
\cite[e.g.][]{bloom02,stanek03}, supporting a physical connection
between the bursts and the catastrophic death of massive stars
\cite{woosley93,paczynski98}.  Studies of the circumburst environment
together with the progenitor stars therefore bear directly on our
understanding of star formation and metal production in the early
universe.

The extreme brightness of GRB optical afterglows, albeit brief, offers
a unique means to unveil the physical conditions of the ambient
medium around the burst progenitors.  For example, low-resolution
afterglow spectra have yielded a few diagnostic measurements such as
the H\,I column density, the gas metallicity, and the dust-to-gas
ratio of the GRB hosts \cite[e.g.][]{sff03,vel04}.  High-resolution
spectroscopy of the optical afterglows has further uncovered detailed
kinematic signatures, population ratios of excited ions, and chemical
compositions of the interstellar medium (ISM) and the circumstellar
medium (CSM) of the progenitor star of the GRB
\cite[e.g.][]{fiore05,chen05,jxp06}.  

The majority of GRB host galaxies are known to have large neutral gas
content with neutral hydrogen column density much beyond $\log\,N({\rm
H\,I})=20.3$ \cite[e.g.][]{vel04}, the standard threshold for
selecting damped \lya\ absorbers (DLAs).  DLAs selected along random
lines of sight toward background quasars dominate the mass density of
neutral gas in the universe and represent a uniform sample of distant
galaxies with known $N(\hI)$.  GRB host DLAs ($z_{\rm DLA}=z_{\rm
GRB}$), which are presumably selected by vigorous star formation and
therefore probe deep into the center regions of distant galaxies,
presents a nice complement to classical DLA system ($z_{\rm DLA}<z_{\rm
GRB}$), which arises preferencially at large galactocentric radii due
to a larger cross-section of the outskirts than the inner regions.  In
this article, we summarize our analysis of the circumstellar medium
observed in two GRB host galaxies, and compare with what is known for
classical DLAs.

\section{Absorption-line Properties of GRB Hosts}

\begin{figure}
  \includegraphics[scale=0.6, angle=270]{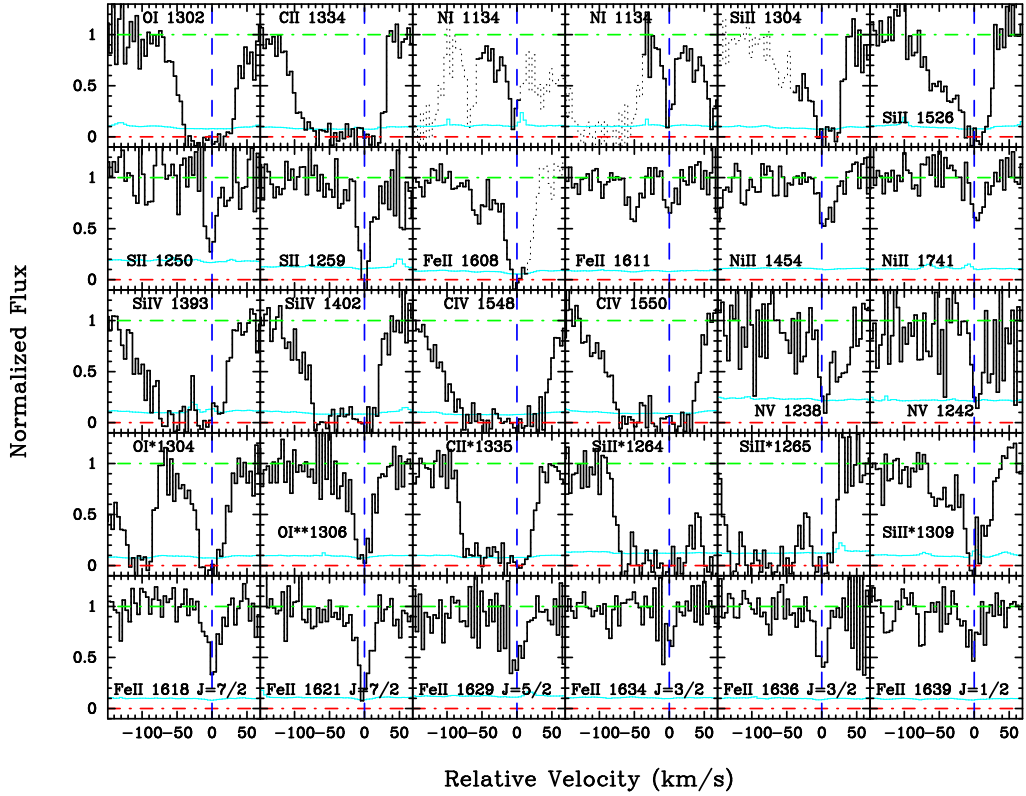}
  \caption{Absorption profiles of various ionic transitions found at
  the host redshift of GRB\,050730.  Resonance transitions are
  presented in the top three rows, while absorption features from
  excited O$^0$, C$^+$, Si$^+$, and Fe$^+$ are presented in the bottom
  two rows.  Dotted curves indicate contaminating features that are
  not related to the marked transitions.  The zero relative velocity
  corresponds to redshift $z=3.96855$.}
\end{figure}

High-resolution echelle spectra of GRB\,050730 were obtained using the
MIKE echelle spectrograph on the Magellan Clay telescope.  The spectra
cover a wavelength range from $\lambda = 3300$ \AA\ to 9000 \AA\ with
a spectral resolution of FWHM $\approx 10$ \kms\ at wavelength
$\lambda=4500$ \AA\ and $\approx 12$ \kms\ at $\lambda=8000$ \AA.  We
have identified a suite of metal-absorption lines at the redshift of
the GRB $z=3.96855$, including neutral species such as O\,I,
low-ionization transitions such as C\,II, Si\,II, S\,II, Ni\,II, and
Fe\,II, and high-ionization transitions such as C\,IV, Si\,IV, and
N\,V.  We have also identified strong fine-structure lines such as
O\,I$^*$, O\,I$^{**}$, Si\,II$^*$, C\,II$^*$, and Fe\,II$^*$.

Absorption profiles of these transitions are presented in Figure 1.
The presence of strong fine-structure transitions indicate an extreme
ISM environment with high gas density that is rarely observed in
classical DLA systems.  Furthermore, the profiles of well resolved
lines (e.g.\ S\,II\,1250) show that $>90$\% of the neutral gas is
confined to a velocity width 20 \kms, which is considerably smaller
than the median value of intervening DLA systems and implies a
quiescent environment.  The asymmetry of these line profiles is also
suggestive of an organized velocity field, e.g.\ rotation or outflow.

Additional properties measured for the host of GRB\,050730 are
summarized as follows.  (1) We measure $\log\,N(\hI)=22.15\pm 0.05$
and $[{\rm S}/{\rm H}]=-2.0\pm 0.1$.  Because S is non-refractory, its
gas-phase abundance gives a direct measurement of the gas metallicity.
(2) We find $[{\rm S}/{\rm Fe}]=+0.3$, consistent with the gas phase
$[\alpha/{\rm Fe}]$ measurements of low-metallicity DLA systems.  Even
if we adopt an intrinsic solar abundance pattern, the dust-to-gas
ratio in the host ISM is very low \cite[c.f.][]{sff03}.  (3) We
measure $[{\rm N}/{\rm S}]=-1.0 \pm 0.2$, again consistent with
low-metallicity DLA systems.  (4) We detect no molecular lines in the
\lya\ forest with a 3-$\sigma$ upper limit to the molecular fraction
$f_{\rm H_2}\equiv 2\,N({\rm H_2}) /[2\,N({\rm H_2})+N(\hI)]<10^{-8}$.
The lack of H$_2$ suggests a warm gas phase, consistent with the
implication from the detections of excited Si$^+$ and Fe$^+$.

\begin{figure}
  \includegraphics[scale=0.65, angle=270]{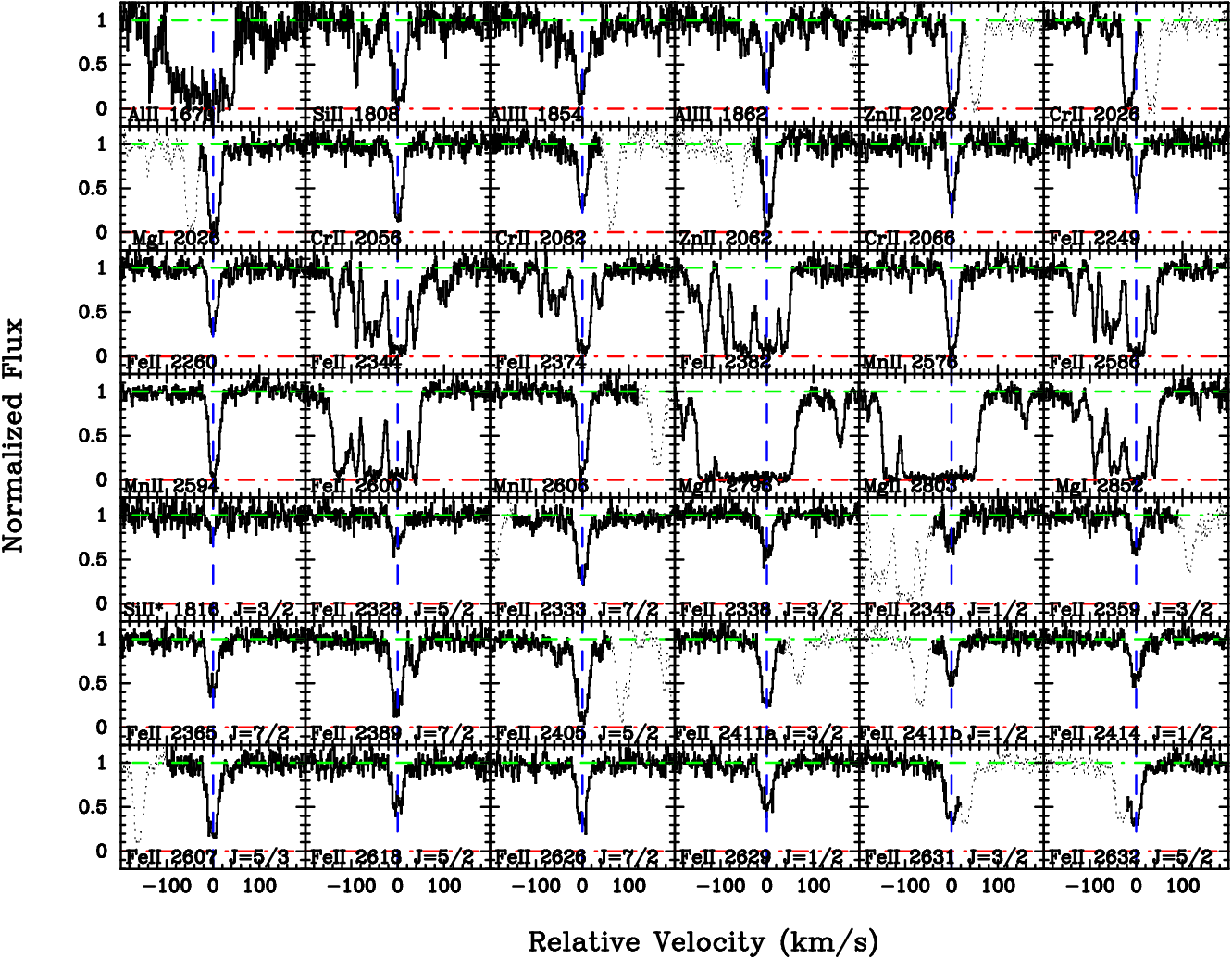}
  \caption{Absorption profiles of various ionic transitions found at
  the host redshift of GRB\,051111.  Resonance transitions are
  presented in the top four rows, while absorption features from
  excited Si$^+$, and Fe$^+$ are presented in the bottom three rows.
  The zero relative velocity corresponds to redshift $z=1.54948$.}
\end{figure}

High-resolution echelle spectra of GRB\,051111 were obtained using the
High Resolution Echelle Spectrometer (HIRES) on the Keck~I telescope.
The spectra cover a wavelength range from $\lambda = 4165 - 8720$\AA\
with a spectral FWHM resolution of $\approx 5 \kms$.  We have 
identified a suite of metal absorption lines at the redshift of
the GRB $z=1.54948$, including resonance transitions such as Si\,I,
Mg\,I, Zn\,II, Cr\,II, Mg\,II, Fe\,II, Mn\,II, and Al\,III, as well
as fine-structure transitions due to excited Si$^+$ and Fe$^+$.

Absorption profiles of these transitions are presented in Figure 2.
We find that while saturated lines such as Fe\,II, Mg\,I and Mg\,II
exhibit multiple components over a velocity range from $-140$ \kms\ to
$40$ \kms, well resolved lines such as Cr\,II and weak Fe\,II show
that $>90$\% of the neutral gas is confined to a velocity width 30
\kms.  In addition, the Si\,II 1808 and Zn\,II\,2026,2062 transitions 
are saturated, indicating $\log\,N({\rm Zn}) > 13.5$ and $\log\,N({\rm
Si}) > 15.9$.  These measurements represent by far the largest Zn and
Si column densities observed in QSO absorption line systems.  The
presence of Fe$^+$ in all four excited $^6D_{J}$ states with $J=7/2,
5/2,3/2$, and $1/2$ suggests a large gas density in the absorbing
medium.  The weak Mg\,I\,2026 transition appears to be saturated,
indicating a warm gas of temperature $T>1000$ K.

The hydrogen \lya\ transition of GRB\,051111 is not observed in the
ground-based data.  Adopt the observed lower limit to the Zn abundance
$\log\,N({\rm Zn}) > 13.5$, we estimate $\log\,N(\hI) >20.8$ for gas
of solar metallicity.  Lower metallicity would lead to higher $N(\hI)$
for the neutral gas content in the host ISM.  In addition, the
observed Zn to Fe ratio is $[{\rm Zn}/{\rm Fe}] > 1.2$.  This large
ratio suggests significant differential depletion in the host of
GRB\,051111, contrary to what is observed for the host of GRB\,050730.

\section{Circumburst Environment of GRB Progenitors}

An emerging feature of GRB progenitor environments is the presence of
strong fine-structure transitions from excited states of C$^+$,
Si$^+$, O$^0$, and Fe$^+$.  In contrast to resonance transitions of
the dominant ions in neutral gas, these absorption lines can reveal
the temperature and density of the gas, as well as the ambient
radiation field.  In particular, detections of Fe\,II fine structure
transitions have only been reported in rare places such as broad
absorption-line (BAL) quasars, $\eta$ Carinae and the circumstellar
disk of $\beta$ Pictoris.  Identifications of strong Fe\,II
fine-structure transitions therefore suggest extreme gas density and
temperature in the GRB progenitor environment.

We examine the excitation mechanism and gas density of the absorbing
medium through comparisons of the observed relative abundances between
different excited states of the Fe$^+$ ion.  The population ratios
between different states are determined based on the balance between
excitation and de-excitation rates.  When the density is
sufficiently large that the collisional de-excitation rates exceed the
spontaneous decay rate, the excited states are populated according to
a Boltzmann distribution,
\begin{equation}
\frac{n_i}{n_j} = \frac{g_i}{g_j} \exp [ -(E_{ij}/k) / 
 T_{Ex} ],
\end{equation}
where $g_i$ is the degeneracy of state $i$, $E_{ij} \equiv E_i - 
E_j$ is the difference in energy between the two states, and
$T_{Ex}$ is the excitation temperature.  

Figure 3 shows the observed column density $N_i$ scaled by the
corresponding degeneracy $g_i$ for each of the Fe$^+$ excited states
as a function of the energy $E_{ij}$ above the ground state $J=9/2$.
The error bars reflect 1-$\sigma$ uncertainty in $N_i$.  The solid
(blue) curve indicates the best-fit Boltzmann function with a best-fit
excitation temperature $T_{\rm Ex}=6100$ K for GRB\,050730 and $T_{\rm
Ex}=2600$ K for GRB\,051111.  The inset shows the minimum $\chi^2$
values as a function of $T_{\rm Ex}$.  The reduced $\chi^2$ is nearly
unity, $\chi_\nu^2=1.02$, supporting our assumption that the Boltzmann
function is a representative model.

\begin{figure}
  \includegraphics[scale=0.27, angle=90]{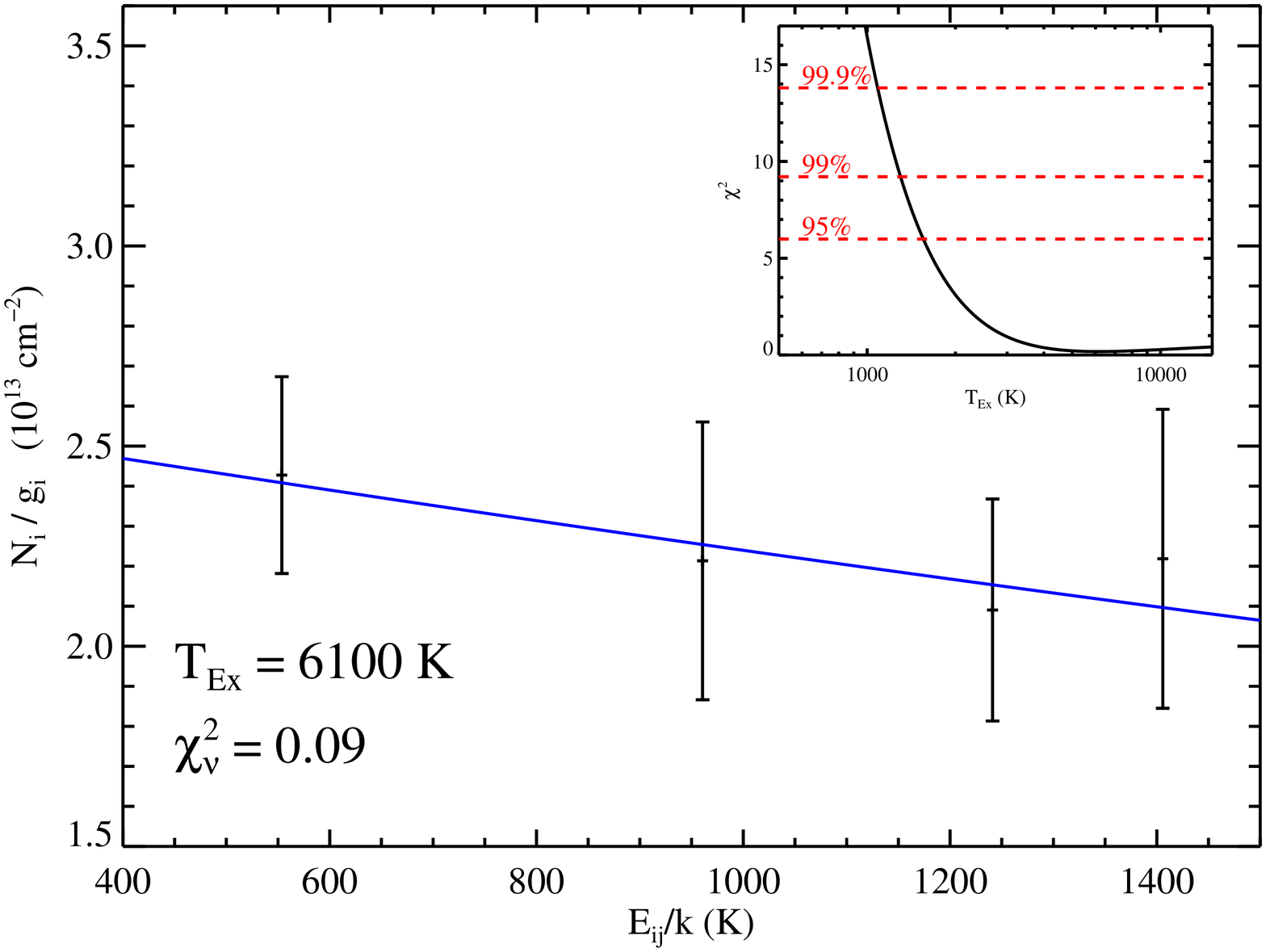}
  \includegraphics[scale=0.27, angle=90]{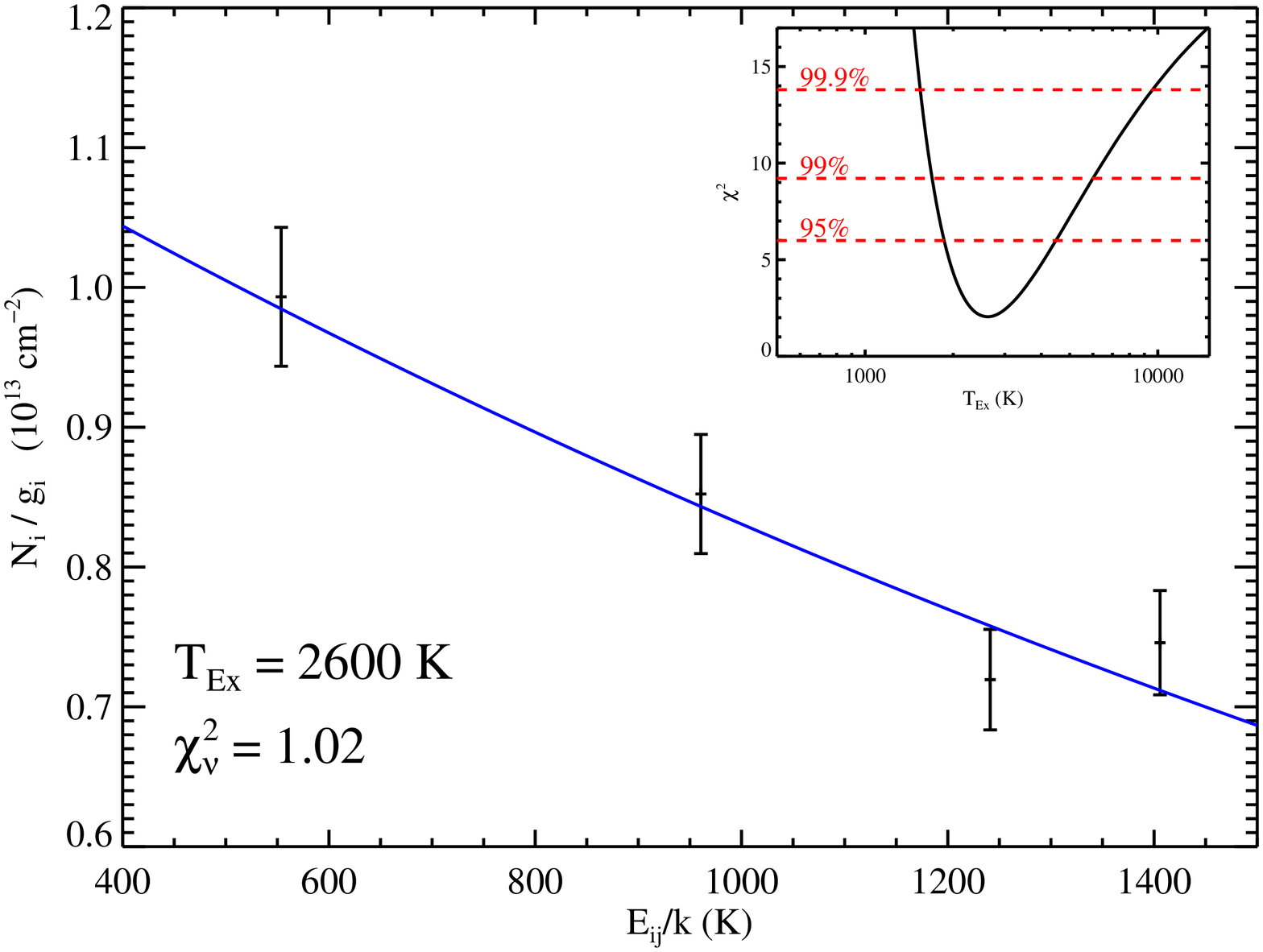}
  \caption{Relative abundance ratios between different excited Fe\,II
  states, $N_i/g_i$ where $g_i$ represents the degeneracy of the
  corresponding level, as a function of their energy $E_{ij}$ above
  the ground state.  We present measurements for the host of
  GRB\,050730 in the left panel and GRB\,051111 in the right panel.
  The solid curve in each panel represents the best-fit function $A
  \exp[-(E_{ij}/k)/T_{Ex}$ for a minimum $\chi^2$ value.  We find a
  best-fit excitation temperature $T_{Ex} = 6100$\,K for GRB\,050730
  and $T_{Ex} = 2600$\,K for GRB\,051111.  The insets show the reduced
  $\chi^2$ values for a range of $T_{Ex}$ values.}
\end{figure}

Adopting the best-fit temperature for the absorbing medium, we can
further constrain the electron density based on the observed column
density ratios.  We calculate the expected population ratio between
excited fine-structure states $i$ and ground state $J=9/2$ for gas of
$T=6100$ K and $T=2600$ K, using the PopRatio software package
\cite{silva02}.  We find that $n_e > 1000 \,{\rm cm}^{-3}$ for both
systems.  To estimate the total gas density $n_{\rm H}$ based on the 
estimated $n_e$ requires knowledge of the ionization state of the 
absorbing gas traced by the excited Fe$^+$.  The large $N(\hI)$
estimated for both systems implies an optical depth to ionizing 
radiation ($h\nu > 1$Ryd) $\tau > 1000$.  While an O or B star (or 
the GRB itself) could significantly ionize a skin-layer on the outside
of the cloud, the majority of the gas would remain neutral.  Therefore, 
the hydrogen gas must be predominantly neutral and we can safely assume 
an ionization fraction $x < 10^{-2}$.  We therefore infer $n_{\rm H} > 
10^5$ cm$^{-3}$ for the absorbing medium, which further leads to an
upper limit to the size of the excited medium $\ell = N(\hI)/n_{\rm H} 
< 10^{17}$\,cm, i.e., less than 1/3 pc.

This small dimension implies that it must arise near the
GRB event (at $r \sim \ell$) or have a nearly unity covering fraction
at large radii (e.g.\ a thin shell).  Additional constraints can be
derived based on absorption features due to neutral 
species and the afterglow light curves.  A detailed discussion is
presented in Prochaska, Chen, \& Bloom (2006).

\section{Classical DLAs versus GRB Hosts}

\begin{figure}
  \includegraphics[scale=0.3]{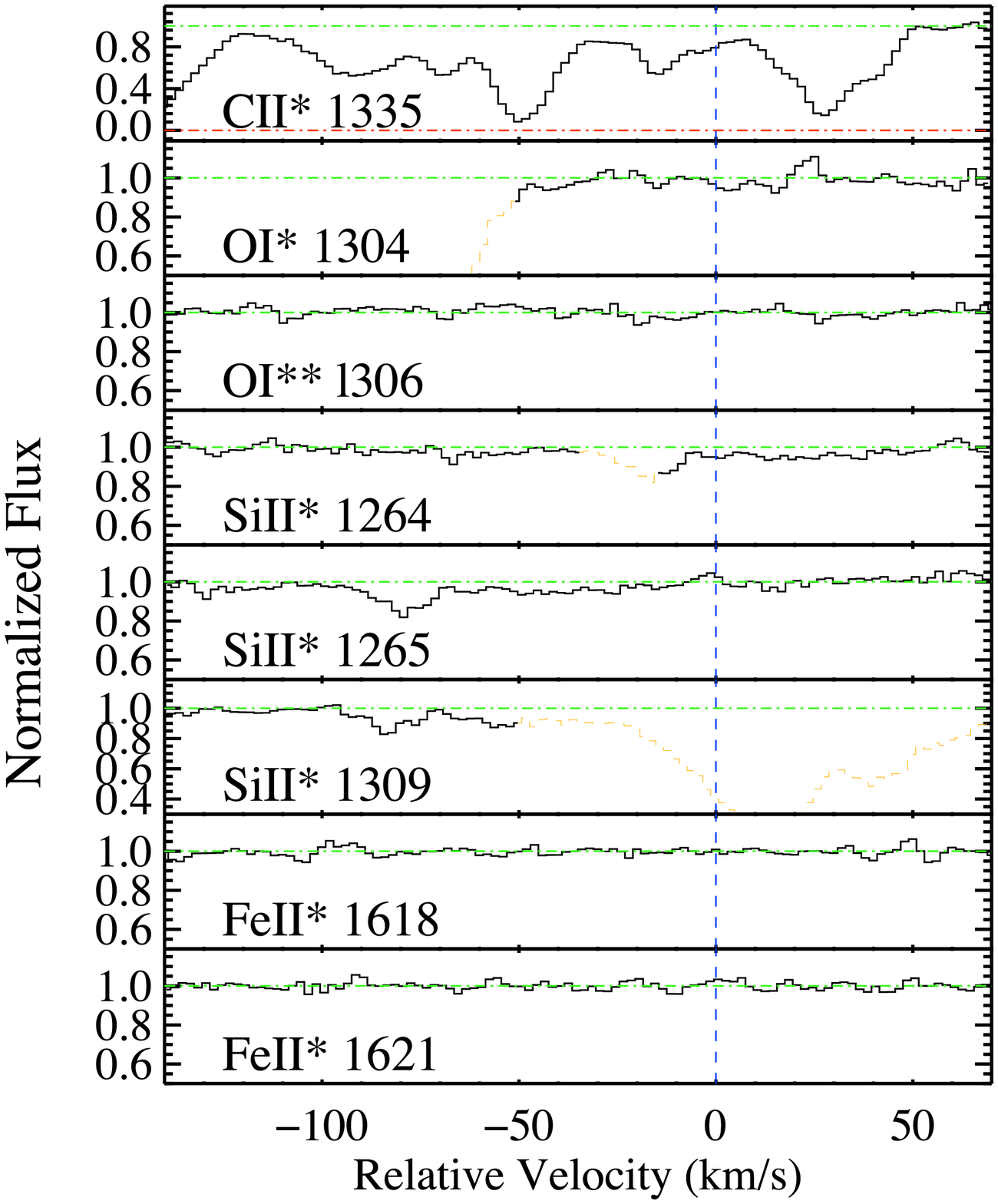}
  \includegraphics[scale=0.3]{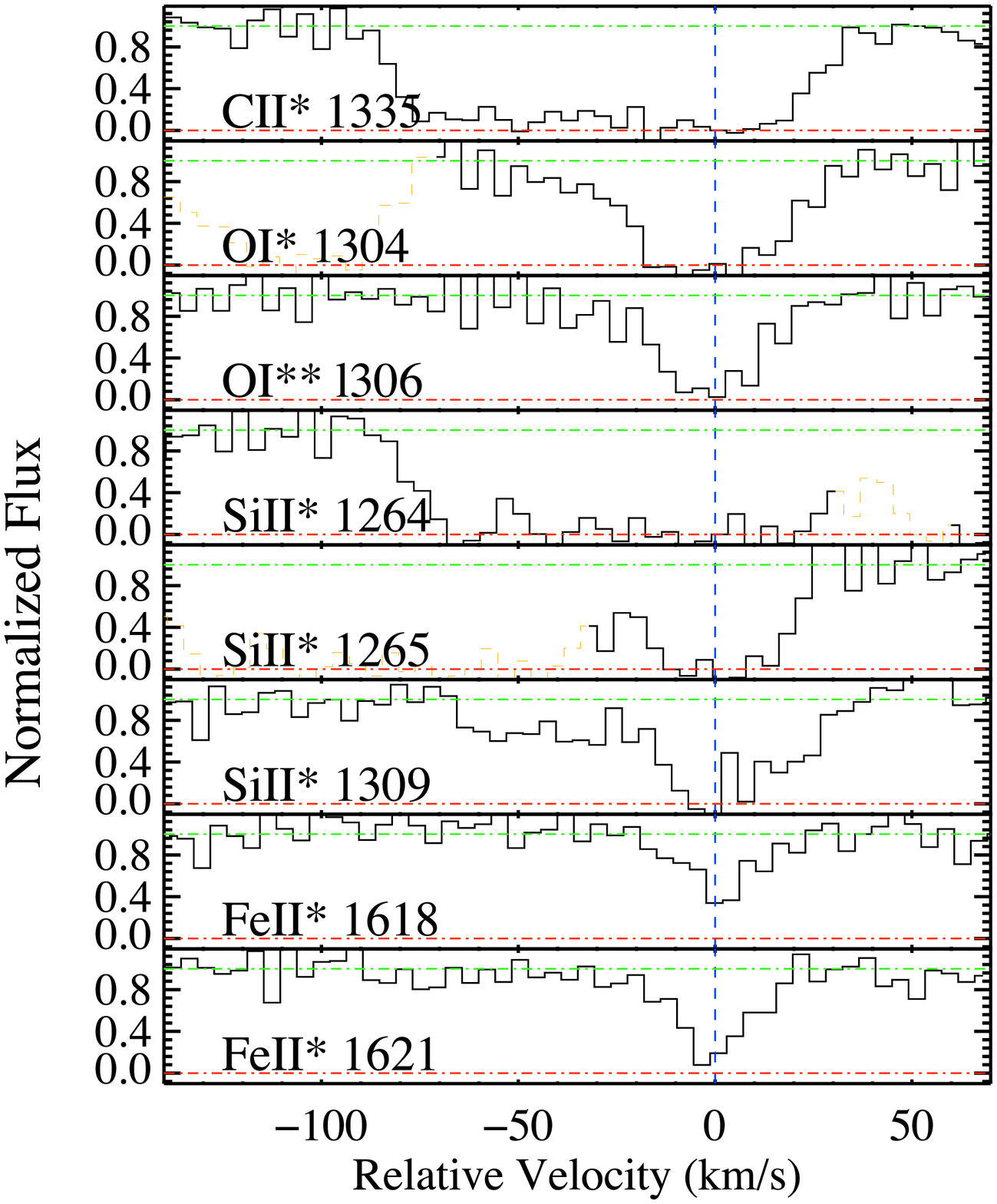} \caption{Comparisons of
  line strengths due to different excited ions observed in
  a classical DLA with $\log\,N(\hI)=21.3$ at $z=2.6264$ (left panel)
  and in the host of GRB\,050730 (right panel).}
\end{figure}

A striking feature is that both GRB hosts exhibit Fe$^+$ and
Si$^+$ fine-structure transitions which have never been observed in
intervening quasar absorption line systems \cite[e.g.][]{howk05}.
Figure 4 presents a sharp contrast in the line strengths of
different excited ions observed in a classical DLA with $\log\,N(\hI)
=21.3$ at $z=2.6264$ and in the host of GRB\,050730.  Our study shows
that aside from having a much higher neutral gas density and
temperature as derived from the observed population ratios between
different excited ions, the GRB hosts have very similar
characteristics to known DLA systems at $z>2$, such as low dust
content, low metallicity, and $\alpha$-element enhanced chemical
composition.

  As bright lighthouses for high-resolution optical spectroscopy,
GRBs are now fulfilling their promise as detailed complementary
probes of the distant universe. It is only a matter of time when high-
resolution spectroscopy at {\bf infrared} wavelengths will afford an
picture of star formation and chemical enrichment in the epoch of
reionization.

%%%%%%%%%%%%%%%%%%%%%%%%%%%%%%%%%%%%%%%%%%%%%%%%
%% BACKMATTER
%%%%%%%%%%%%%%%%%%%%%%%%%%%%%%%%%%%%%%%%%%%%%%%%

\begin{theacknowledgments}
The authors are grateful to Ian Thompson for obtaining the Magellan/MIKE data 
of GRB\,050730.  We thank Grant Hill, Derek Fox and Barbara Schaefer for their 
roles in obtaining the Keck/HIRES data of GRB\,050111.  H.-W.C., J.X.P., and 
J.S.B.\ are partially supported by NASA/Swift grant NNG05GF55G.
\end{theacknowledgments}




%\bibliography{sample}

%  \typeout{******************************************}
%  \typeout{** Please run "bibtex \jobname" to optain}
%  \typeout{** the bibliography and then re-run LaTeX}
%  \typeout{** twice to fix the references!}
%  \typeout{******************************************}

\end{document}